\documentclass[twocolumn,prl]{revtex4}
\RequirePackage{xspace}
\usepackage{graphicx}

\def\be{\begin{equation}}
\def\ee{\end{equation}}
\def\bearr{\begin{eqnarray}}
\def\eearr{\end{eqnarray}}
\def\hlf{${\frac{1}{2}}$~}
\begin{document}

\title{$\frac{3}{2}$-Fermi liquid: the secret of high-Tc cuprates$^*$}

\author{ G. Baskaran\email}
\affiliation{Institute of Mathematical Sciences, C.I.T. Campus, Chennai 600 113, India}

\begin{abstract}
Electrical and magnetic properties of underdoped cuprates in the spin gap phase, a precursor to high Tc superconducting state, is riddled with puzzles. We propose a novel reference state to study this phase, where each hole dopant adds one real holon and one real spinon, with Haldane exclusion statistics $g_{\rm h} =1$ and $g_{\rm s} = \frac{1}{2}$, to a fairly inert (pseudo gaped) spin liquid vacuum, resulting in a low density spin-charge liquid. Spins and charges interact and form a novel collective state, a \textit{$\frac{3}{2}$-Fermi liquid}. A holon pairs with a spinon; this hole like fermion composite carries charge +e and a spin-\hlf moment and novel exclusion statistics $g_{\rm hole} = \frac{3}{2}$. We explain an anomalous expansion of Fermi sea area by $\frac{3}{2}$, seen in recent quantum oscillation experiments at two different dopings.
\end{abstract}

\maketitle
From early days of quantum theory of solids, `ideal' reference systems such as a harmonic solid, 
ideal Bose gas, ideal Fermi gas have played a fundamental role in shaping our understanding of real and complex solids. For example, BCS theory for real superconductors like Hg or Pb, is built on a Fermi liquid reference state and superconductivity is viewed as an instability of an ideal Fermi gas in the presence of weak attractive interaction. With the advent of high Tc superconductivity in cuprates\cite{bednorz} and further developments, a need for new reference states has become clear. The theory of resonating valence bond state (RVB) in doped Mott insulators\cite{pwaScience,pwaBook,bza,Kivelson,pwa1} has provided a deep understanding of the mechanism of high Tc superconductivity and properties of the superconducting state. However, in spite of several valiant attempts and significant insights in the last 2 decades
\cite{pwaSpinLock,vanila,gbIran,leeRMP}, a satisfactory and simple `ideal gas like' reference system, that describes the rich physics of spingap phase and its neighbourhood, is lacking. 

A better understanding of the spingap phase is a need of the time, as there are precise and new experimental results\cite{qOsc1,qOsc2,OngNernst} that demands explanation. We also need to understand how high Tc superconductivity emerges from this phase and how this phase crosses over to other exotic normal phases such as the high temperature incoherent metal and over doped non-Fermi liquid phase. We propose a new reference state to characterise the spin gap phase. Our theory is inspired by i) Anderson's recent proposal of spin-charge locking\cite{pwaSpinLock} in the spin gap phase, ii) an early notion of exclusion statistics by Haldane\cite{haldane} and iii) very recent quantum oscillation measurements in underdoped cuprates\cite{qOsc1,qOsc2}, presenting a remarkable evidence for quasi particles and Fermi surface pockets, with an area that is $\frac{3}{2}$ times larger than expected based on hole counting
and other measurements ! There has been some theoretical attempts\cite{zhangRice07} to understand this areal anomaly.

The quantum spin liquid or RVB state is a generic name for a family of doped and undoped Mott insulating states. The present work shows that the RVB state relevant for the spingap phase is a new state to be called \textit{$\frac{3}{2}$-Fermi liquid} with properties that are different from Fermi liquids and Luttinger liquids in some fundamental way. Further \textit{$\frac{3}{2}$-Fermi liquid is not adiabatically connected to the reference non interacting electron system}.  Luttinger theorem is strongly violated. After formulating our theory we show how it explains an anomalous Fermi sea area seen in recent quantum oscillation measurements.
\begin{figure}
\includegraphics*[width=3.5cm]{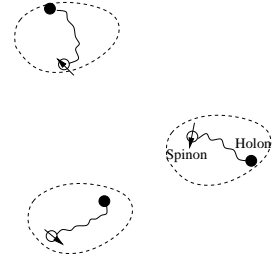}
\caption{\label{fig1}
A low density liquid of paired spinon-holon composite. The pairing may be thought to arise from
dynamically generated RVB gauge fields.}
\end{figure}

We wish to study a lightly hole doped spin-$\frac{1}{2}$ Mott insulator. When doping fraction is x $<<$ 1, upper Hubbard band excitations continue to have a finite energy gap. Consequently superexchange survives and charge dynamics arises only from the hole motion. The t-J model plus some unscreened short range coulomb repulsion describes this `projected metal'. 

Our theory of $\frac{3}{2}$-Fermi liquid starts with the following hypothesis: there exists a reference Mott insulating spin liquid phase, which has no holons and negligibly small density of thermal spinons (pseudo gap) in the range of temperature of interest to us. Each doped hole adds one holon and one spinon to this fairly inert spin liquid. What we have is a fluid of interacting doped holons and equal number of doped spinons, which we call as a \textit{spin-charge fluid}. We do not go into the nature and origin of this interaction in the present paper.  Following Anderson\cite{pwaSpinLock}, we will take it as a phenomenological fact that as one enters the spin gap regime, from high temperature side, there is a spin-charge locking. In our theory, we view it as pairing of one doped holon and one doped spinon and condensation of this spinon-holon (fermion) composite into a novel Fermi sea. Total number of holons is conserved because of charge conservation. However, doped spinon number is not a conserved quantity in general; in fact, we expect them to form spin singlets and disappear into the vacuum. Our assumption is that \textit{spin-charge locking and coulomb repulsion between the composite keeps the spins of the doped spinons nearly free and uncorrelated} (figure 1).

Spinon and holon of RVB theory are topological excitations\cite{Kivelson,pwa1}; this property manifests an underlying topological order\cite{Kivelson}. Their addition or removal leads to a shakeup of the many body states they live, in a global fashion; for example, in 1D repulsive Hubbard model these shakeups appear as self consistent change of all the pseudo momenta of spins and charges. \textit{We suggest that exclusion statistics, introduced by Haldane\cite{haldane}, is a key and minimal notion that needs to be incorporated, in order to solve our complex spin gap problem}, where we do not have the advantage of exact solutions or Bethe type ansatz. To characterise the spin-charge fluid we need to find the exclusion statistics of the constituent spinon ($g_{\rm s}$) and holon ($g_{\rm h}$) separately first and then that of the spinon-holon composite ($g_{\rm hole}$). We find that exclusion statistics of the spinon-holon pair is a sum of their individual exclusion statistics, $g_{\rm hole} = g_{\rm s} + g_{\rm h}$.

In Haldane's theory of exclusion statistics, a reduction in Hilbert space dimension of a given single quasi particle, due to presence of other quasi particles, determines exclusion statistics parameters $g$. Let $\Psi (r_1,..r_i,..r_{M})$ be the wave function of a M identical quasi particle system in coordinate representation. For our purpose we consider a lattice system containing N lattice sites. We leave free the i-th quasi particle coordinate and freeze coordinates of the rest M-1 quasi particles at a typical configuration denoted by $C$ and expand $\Psi$ in a set of single particle basis wave function of the i-th particle $\{\phi_\nu^{C}(r_i)\}$ :
\be
\Psi (r_1,..r_i,..r_{M}) = \sum_{\nu = 1,..Q} a_{\nu}^{C} \phi_{\nu}^{C}(r_i)
\ee
The set of single particle wave functions $\{\phi_\nu^{C}(r_i)\}$ span a one particle Hilbert space with dimension Q. For exclusion statistics to become meaningful, Q should be independent of the configuration $C$. In general Q will change as particles are added. Haldane defines the exclusion statistics parameter through a suggestive differential relation $\Delta Q \equiv - g ~\Delta M$. It is remarkable that in correlated (non slater determinantal) states like Laughlin quasi particle states in fractional quantum Hall effect or spinon states of RVB theory $g$ can be fractional; this is due to an intrinsic non-orthogonality of the `topological excitations', which is transparent in real space basis.  How exclusion statistics, gets translated into k-space is a difficult issue, as we will see below. 

For holon it is easily shown that $g_{\rm h}$ = 1. It follows from the hard core and band filling nature. To find the exclusion statistics of spinons, first we will offer a k-space analysis
and find number of allowed single spinon states in the Brillouin zone (BZ) of a spin liquid state. One dimensional Haldane-Shastry model\cite{HaldaneShastry} is known to exhibit a remarkable property that the number of single spinon states in k-space is only half the number of k-points in the BZ. This implies an exclusion statistics of $g_{\rm s} = $ \hlf for a spinon. We will show, using variational (Fermionic) RVB wave functions for spinon states that \textit{allowed k-values of a single spinon is restricted to only half the number of points in BZ for RVB states in any dimension}, implying that the mutual exclusion statistics of a spinon is \hlf, independent of their spin states.  
\begin{figure}
\includegraphics*[width=3.5cm]{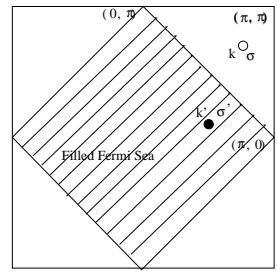}
\caption{\label{fig2}
Two spinon states created by a Gutzwiller projected particle-hole excitation in a Fermi sea of
half filled band of free electrons in a 2D square lattice. By construction, momentum of a given spinon is limited one half of the BZ (either inside or outside the Fermi sea, which get identified
after GW projection).} 
\end{figure}

In RVB theory\cite{pwaScience,bza}, Gutzwiller projection and mean field states (Slater determinants) play a fundamental role. Let us start with a simple mean field solution, a half filled band of free Fermi sea (zero-flux solution), $|FS\rangle$ in 2 dimensions. A two spinon state\cite{bza}, with momentum and spin projections $({\bf k},\sigma)$ and $({\bf k}',\sigma')$ is defined as (figure 2):
\be
|{\bf k} \sigma,{\bf k}' \sigma'\rangle \equiv P_{\rm G} c^{\dagger}_{{\bf k} \sigma}c^{}_{{\bf k}'{\bar{\sigma}}'}
|FS\rangle
\ee
It is easy to see that the above construction creates two unpaired spins with spin projections 
$\sigma$ and $\sigma'$, in the background of resonating valence bonds and maintaining single electron occupancy at every site. As two spinons are created from a half filled band Fermi sea containing 
a particle-hole pair, the allowed momenta of the two spinons ${\bf k}$ and ${\bf k}'$ are the two 
halves of the BZ (figure 1). Further, it is easy to show using equation (2) that 
\be
|{\bf k} \sigma,{\bf k}' \sigma'\rangle \equiv |{\bf k} + {\bf Q}~ \sigma,{\bf k}'+ {\bf Q}~ \sigma'\rangle
\ee
where ${\bf Q} \equiv (\pi,\pi)$. The above completes the proof that the spinon band covers only half the BZ. That is, there is one independent spinon state per two points in k-space.
This proof goes through, as long as the starting mean field RVB state is a Slater determinant that fills any half of mean field single particle states with up and down spin electrons. 

One can get a feel for the above state reduction, using Haldane's argument, which uses non-orthogonality of site localised basis for spinons. For an insulating  RVB state containing N spins, N$_{sp}$ spinons (unpaired spins), and $\frac{N-N_s}{2}$ unbroken singlet bonds on a lattice, the Hilbert space dimension of a single spinon is $1 + \frac{N-N_s}{2}$, independent of the spin projection of the spinon. This is understood as follows. A given spinon can occupy the site that it is initially on or it can be moved to any site that is part of a pair. But if $|1\{23\}\rangle$ represents a three-site wave function where site 1 is unpaired and sites 2 and 3 are paired, nonorthogonality means that $|1\{23\}\rangle = \frac{1}{{\sqrt2}}|2\{13\}\rangle - |3\{12\}\rangle $. Hence only the combination $|2\{13\}\rangle + |3\{12\}\rangle $ is independent of $|1\{23\}\rangle$. There is thus one extra independent spinon state per bond (half the number of sites the bonds occupy). This makes $g_{
m s} = \frac{1}{2}$, independent of their spin states.

What is the exclusion statistics of a spinon-holon composite ? We answer this question by a heuristic argument. Fractional exclusion property is a manifestation of real space non-orthogonality of the quasi particles. The spinon component of the composite continues to have the same non-orthogonality property for every spinon-holon configuration in a composite. Thus it follows that mutual exclusion statistics of a composite is additive and independent of the spin states of the two fermions: i.e., $g_{\rm hole} =   g_{\rm s} + g_{\rm h} = \frac{1}{2} + 1 = \frac{3}{2}$. That is, as we add a spinon-holon composite to the system, its holon component excludes one state and spinon component excludes half a state for the next composite that enters. 

So far our discussion has been general. Now we use specific spin liquid state relevant for 2D cuprates and get information about the location and volume Fermi sea pockets, that a collection of fermions with a $\frac{3}{2}$ exclusion statistics will condense into. We start with the $\pi$-flux (s+id) RVB mean field solution\cite{affleckMarston} of the spin-\hlf Heisenberg antiferromagnetic Hamiltonian $H = J\sum_{\langle ij \rangle} \textbf{S}_i\textbf{S}_j$, on  a square lattice with a nearest neighbour exchange J. This variational solution, after Gutzwiller projection, is known to give a good ground state energy. It has nodal spinon excitations at $(\pm {\frac{\pi}{2}},\pm {\frac{\pi}{2}})$; that is, after linearisation, a  Dirac cone like spectrum is obtained: $\epsilon_{\rm sp}(q) \approx \hbar v_{{\rm sp}} q$. Here $v_{\rm{sp}} \sim Ja $ is the spinon Fermi velocity, a is the lattice constant and q is measured from the four nodes. The linearly vanishing density of states of spinons as its energy tends to zero ensures that we have a pseudo gap and small number thermal spinons at low temperatures. The massive holon excitation occurs at the $ (0,0)$ point in the BZ (fig 1). It has a quadratic dispersion, $\epsilon_h \approx \frac{\hbar^2 q^2}{2m^*}$. Here m$^* \sim \frac{\hbar^2}{t a^2}$ is the holon effective mass. 

As discussed earlier, each added hole contributes one real spinon and one real holon in their respective low energy regions in k-space. In the slave boson mean field theory, without local constraints, holons and spinons are non interacting: doped holons Bose condense at the $\Gamma$ point and spinons build four identical Fermi pockets at the $(\pm {\frac{\pi}{2}},\pm {\frac{\pi}{2}})$ points (Figure 3a). In the RVB gauge theory\cite{gaugeGB,wiegman,leeRMP}, which takes into account fluctuations arising from double occupancy constraints, dynamically generated RVB gauge fields appear. In our picture, we have matter (spinons and holons), interacting through residual dynamically generated gauge fields. We view the locking phenomenon as formation of a weak, spinon-holon bound state, induced by dynamically generated gauge fluxes (figure 1). We take the spinon-holon pairing as a phenomenological fact and do not go into details in the present paper. As spinons and holons have their energy minima at $(\pm {\frac{\pi}{2}}, \pm {\frac{\pi}{2}})$ and $(0,0)$ in k-space, we find that the paired spinon-holon composites have their energy minima at the finite momenta, $(\pm {\frac{\pi}{2}}, \pm {\frac{\pi}{2}})$. 
\begin{figure}
\includegraphics*[width=7.0cm]{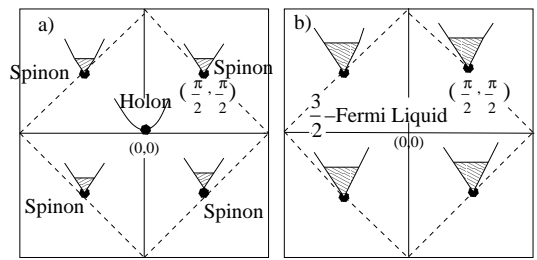}
\caption{\label{fig1}
a) spinon and holon spectrum in the $\pi$-flux (s+id) RVB mean field solution. In a slave boson mean field theory, the doped spinons occupy and form a spinon Fermi sea and the doped holons Bose condense at the $\Gamma$ point. b) Our proposal: the spectrum of the spinon-holon paired composite and the $\frac{3}{2}$-Fermi liquid they form in k-space.}
\end{figure}

Since there are four minima for our $\frac{3}{2}$ fermions in k-space, spinon-holon composite created by doping will build four Fermi pockets around these equivalent points in k-space (figure 3b). As we mentioned earlier, the mutual exclusion statistics of $\frac{3}{2}$ is independent of the spin state of the holes. It means that our more exclusive fermion occupies an average of $\frac{3}{2}$ k-points, below the Fermi level. We incorporate this in the following fashion: each low energy k-state, below a Fermi momentum is occupied by our fermion with an average occupancy of $\frac{2}{3}$, independent of the spin state. This leads to an average number of $\frac{3}{2}$ k-state per fermion. Further, for simplicity, we assume that the spins are dangling and form a paramagnetic gas in k-space. Knowing the technical difficulties in constructing the Bethe ansatz in 1D systems, it is clear that it will be rather hard to give a wave function representation of our $\frac{3}{2}$ Fermi sea, in terms of underlying electron coordinates. However, a realization that such a fractional occupancy and novel Fermi sea is possible, is a key step for a better understanding of the spin gap phase.

Each Fermi sea around the $(\pm {\frac{\pi}{2}},\pm {\frac{\pi}{2}})$ points will include
$\frac{3}{2} \frac{xN}{4}$ k-points below its Fermi surface. In other words the Fermi surface area of each hole pocket is  $\frac{3}{2} \frac{xN}{4}$. As each k-point below the Fermi sea has an average occupancy of $\frac{2}{3}$, we get the total number of holes as $xN$. 

Thus we see that \textit{the new `hole' quasi particle, the weakly bound holon and spinon has no adiabatic connection to `hole' of the reference non interacting system having a total of N(1-x) electrons, even though they both carry charge +e and spin-\hlf moment.}. 

To understand nature of spingap phase, devoid of complications arising from disorder, high quality single crystals of underdoped $YBa_2Cu_3O_{6.5}$ \cite{qOsc1}and two chain compound $YBa_2Cu_4O_8$ \cite{qOsc2}have been studied recently. These two underdoped compounds have different doping,
$x \approx 0.1$ and $0.125$ respectively. In these studies a superconducting state is destroyed by a large external magnetic field, thereby accessing a spin gap normal state at sufficiently low temperatures, where inelastic mean free path of the quasi particles become sufficiently large, as to be able to perform complete cyclotron motions. The quantum oscillation measurements in underdoped YBCO revealed presence of 4 closed Fermi surface pockets in the BZ. A theoretical fit, using Fermi liquid formula, gives each hole pocket an area that is $ \approx 1.5 \frac{xN}{4}$. As observed by the authors of the experiments, for both the compounds this area is nearly 1.5 times the value that one will get assuming that each added hole occupies one k-point inside the Fermi sea. However, it is in excellent agreement with our formula Fermi pocket volume of $\frac{3}{2}\frac{xN}{4}$. 

Thus we believe that we have given a microscopic explanation for the observed anomalous Fermi sea volume. It will be very important to confirm our theory and measure the quasi particle Fermi surface shape and volume at low temperature regime, after destroying superconductivity, using similar systems where quantum oscillation measurements have been performed. Existing ARPES results\cite{arpes1,arpes2} in the spin gap phase are above the superconducting temperatures of the under doped regime. In this region, a strong buildup of superconducting amplitude and associated vortex liquid phenomena are likely to destroy the Fermi surface of our reference $\frac{3}{2}$ Fermi liquid.

Having given a novel reference state to describe the low temperature spin gap phase, important tasks remain: i) how superconducting instability arises in the $\frac{3}{2}$ Fermi liquid
at zero temperatures, as we increase doping ii) nature of a suggested vortex liquid above superconducting Tc in the spin gap phase. iii) nature of the quantum phase transition to the over doped regime, where ARPES sees a large Fermi surface, as given by band structure results. 

Constituents of $\frac{3}{2}$-fermion have conflicts, including tendency for spinon pair annihilation, implying different instabilities. It is likely that at T=0, as we increase the doping, superconducting instability arises beyond a critical doping, by annihilation of doped spinons, and a consequent collapse of our spin-charge fluid into a charged fluid. In the same way, at finite T in the spin gap phase, the $\frac{3}{2}$ Fermi sea is likely to loose part of the doped spinons as singlet pairs and build a strong superconducting amplitude. Some finite temperature effects can be captured by an `ideal gas' like Haldane-Wu\cite{wu} distribution function of our $\frac{3}{2}$ Fermi sea. We hope to address the above issues in future.

${}^*$ As this preprint was getting completed, a preprint by P.W. Anderson [arXiv:0709.0656] has appeared, with a spirit similar to mine. Title of Anderson's paper inspired me to replace a old title 
by the present one.

\end{document}